# Superconductivity in $FeTe_{1-x}S_x$ induced by electrochemical reaction using ionic liquid solution

Aichi Yamashita[1,2], Satoshi Demura[1,2], Masashi Tanaka[1], Keita Deguchi[1,2], Takuma Yamaki[1,2], Hiroshi Hara[1,2], Kouji Suzuki[1,2], Yunchao Zhang[1,2], Saleem J. Denholme[1], Hiroyuki Okazaki[1], Masaya Fujioka[1], Takahide Yamaguchi[1], Hiroyuki Takeya[1], Yoshihiko Takano[1,2]

[1]National Institute for Materials Science, 1-2-1 Sengen, Tsukuba, Ibaraki 305-0047, Japan

[2] Graduate School of Pure and Applied Sciences, University of Tsukuba, 1-1-1 Tennodai, Tsukuba 305-8571, Japan

## Abstract

Superconductivity in $FeTe_{0.8}S_{0.2}$ is successfully induced by an electrochemical reaction using an ionic liquid solution. A clear correlation between the Fe concentration in the solution and the manifestation of superconductivity was confirmed, suggesting that superconductivity was induced by the deintercalation of excess iron.

## 1. Introduction

Discovery of the iron-based superconductors led to a wealth of novel superconductors. The iron-based superconductors, such as $LnFeAsO_{1-x}F$ ($Ln$ = Lanthanoid elements), $Ae_{1-x}K_xFe_2As_2$ ($Ae$ = Ca, Sr, Ba), $A$FeAs ($A$ = Li, Na) have a layered structure which consist of blocking and superconducting layers alternately stacked. On the other hand, a particular family of the iron chalcogenides superconductors (termed the 11-system), for instance FeSe, Fe(Te,S), and Fe(Te,Se), have the simplest crystal structure among the iron-based superconductors[1-4]. The iron chalcogenides have only superconducting layers, without blocking layers [5, 6].

However, several studies have found that a small amount of Fe is trapped between the layers during the synthesis [7]. This excess Fe suppresses superconductivity by doping the excess electrons to the superconducting layers [8, 9].

So far, we have been succeeded in inducing superconductivity in the 11-system compound by removing excess Fe from the interlayer by annealing in an organic acid solution, for instance in a citric acid solution [10-13]. But this reaction progresses slowly and is less tunable to induce superconductivity. More sophisticated methods to remove excess Fe are required. Recently, we have reported that to induce superconductivity excess Fe in $FeTe_{0.8}S_{0.2}$ sample was electrochemically removed, (i.e. deintercalated), by applying voltage to the citric acid solution. This is the first report of inducing superconductivity using an electrochemical reaction in the iron-based superconductors [14].

However, since the solution is composed of citric acid and water, inevitably with increasing applied voltage, the electrolysis of water occurs, resulting in the evolution of $O_2$ gas. As a result this produces an oxide film on the sample surface which interferes with the electrochemical reaction. Thus, the change to a water free solution which has good conductivity is required to avoid the electrolysis of water. For such a solution, an ionic liquid is a possible candidate. The ionic liquid does not contain water, suggesting that the electrolysis of water does not occur. On this basis, we selected the ionic liquid (1-Butyl-3-methylimidazolium tetrafluoroborate).

In this paper, we report the results of an electrochemical reaction using this ionic liquid solution and successfully inducing superconductivity.

## 2. Experimental details

Polycrystalline samples of $FeTe_{0.8}S_{0.2}$ were synthesized by a solid state reaction same as previously reported [14]. A mixture of Te (Kojundo Chemical Lab. Co., Ltd., 99.999 %) and S (Kojundo Chemical Lab. Co., Ltd., 99.9999 %) grains, and Fe (Kojundo Chemical Lab. Co., Ltd., 99.9 %) powder with a nominal composition of $FeTe_{0.8}S_{0.2}$ was put into an evacuated quartz tube. After sintering at 600 °C for 10 h, the obtained materials were thoroughly ground and pressed into rectangular-shaped pellets. The pellets were put into the evacuated quartz tube, and then heated at 600 °C for 10 h again.

The electrochemical reaction was performed by a three electrodes method. An Ag plate was adopted as a reference electrode. A Pt plate was used for a counter electrode and the rectangular-shaped sample was used as a working electrode.

The ionic liquid (1-Butyl-3-methylimidazolium tetrafluoroborate) was used as a solution for the electrochemical reaction. The sample, counter and reference electrodes were soaked into a preheated solution at 80 °C. The voltage was applied by using a chronoamperometry for 1 h at 0.5, 0.8, 1.0, 1.1, 1.2, 1.3, 1.4 and 1.5 V (against Ag reference electrode) to each samples, respectively. Note that bubbles which indicate the electrolysis of water were not observed at any of the applied voltages.

Powder X-ray diffraction (XRD) patterns were measured by Rigaku model MiniFlex600 using Cu Kα radiation. The temperature dependence of magnetic susceptibility was measured using a SQUID magnetometer with both zero-field cooling (ZFC) and field cooling (FC) mode from 2.0 to 15.0 K under a magnetic field of 10 Oe. Fe and Te concentration in the solution after reaction was analyzed by using inductively-coupled plasma (ICP) spectroscopy.

Sample surface was observed by Scanning Electron Microscope (SEM) and analyzed by Energy dispersive X-ray spectrometry (EDX).

**3. Results and discussion**

Figure 1 shows the XRD patterns of $FeTe_{0.8}S_{0.2}$ samples of before (as-grown) and after the electrochemical reaction. For the as-grown sample, the obtained patterns are in good agreement with previous reports [14]. The XRD patterns after electrochemical reaction show almost no evidence of change to the sample up to 1.2 V.

Figure 2 shows the temperature dependence of magnetic susceptibility for the as-grown sample and the samples after electrochemical reaction. No diamagnetic signal was observed in the as-grown sample, and we also confirmed that superconductivity was not observed by just soaking the sample in the ionic liquid solution at 80 °C without applying voltage. On the other hand, all the samples after electrochemical reaction show a large diamagnetic signal below 7 K in ZFC mode, corresponding to superconductivity. These results indicate that superconductivity was induced electrochemically.

Each solution after the electrochemical reaction was analyzed using ICP spectroscopy. The Fe and Te concentrations as a percentage were estimated from the detected amount of Fe and Te. The concentration was defined by the following formula,

$$\text{Fe or Te concentration } (\%) = \frac{\text{Detected Fe or Te in the solution (mol)}}{\text{Total Fe or Te of the sample soaked in the solution (mol)}}$$

and then the concentration was plotted in figure 3 using a logarithmic scale as a function of the applied voltage with a shielding volume fraction, which was estimated from the value of ZFC mode at 2 K. The Fe concentration gradually increased with increasing applied voltage, and it shows a bell-shaped curve which has a peak at 1.1 V. The shielding volume fraction also shows the bell-shaped curve, and it shows a similar behavior with the Fe concentration. This result further indicates a clear correlation between the deintercalation of excess Fe and the manifestation of superconductivity. This result suggests that superconductivity was induced by the deintercalation of excess

Fe via the electrochemical reaction. Above 1.3 V, the Fe and Te concentrations sharply increased. At the same time, the shielding volume fraction decreased, suggesting that the superconducting layer has started to decompose around these voltages.

Figure 4 shows the Scanning Electron Microscope (SEM) images and Energy dispersive X-ray spectrometry (EDX) elementary mapping of $FeTe_{0.8}S_{0.2}$ sample after the electrochemical reaction at 1.5 V. The elemental mapping analysis showed that iron and fluorine are mainly deposited on the sample surface. This result indicates that fluorine in the ionic liquid combine with iron of the sample. This impurity deposition is clearly found from 1.3 to 1.5 V.

## 4. Conclusion

We are able to induce superconductivity in $FeTe_{0.8}S_{0.2}$ by a deintercalation of excess Fe via an electrochemical reaction using an ionic liquid solution. The ICP spectroscopy analysis of the solution showed that the Fe ion was effectively deintercalated from the interlayer of sample. The superconducting properties were improved increasing applied voltage, and it shows a bell-shaped curve which has a peak at 1.1 V.

Although the electrolysis of water was not observed around the electrodes,

there is concern about reaction with iron and fluorine. For future work, an ionic liquid which does not react with iron is anticipated as an ideal solution for this method.

**Acknowledgments**

This work was partly supported by the Strategic International Collaborative Research Program (SICORP-EU-Japan).

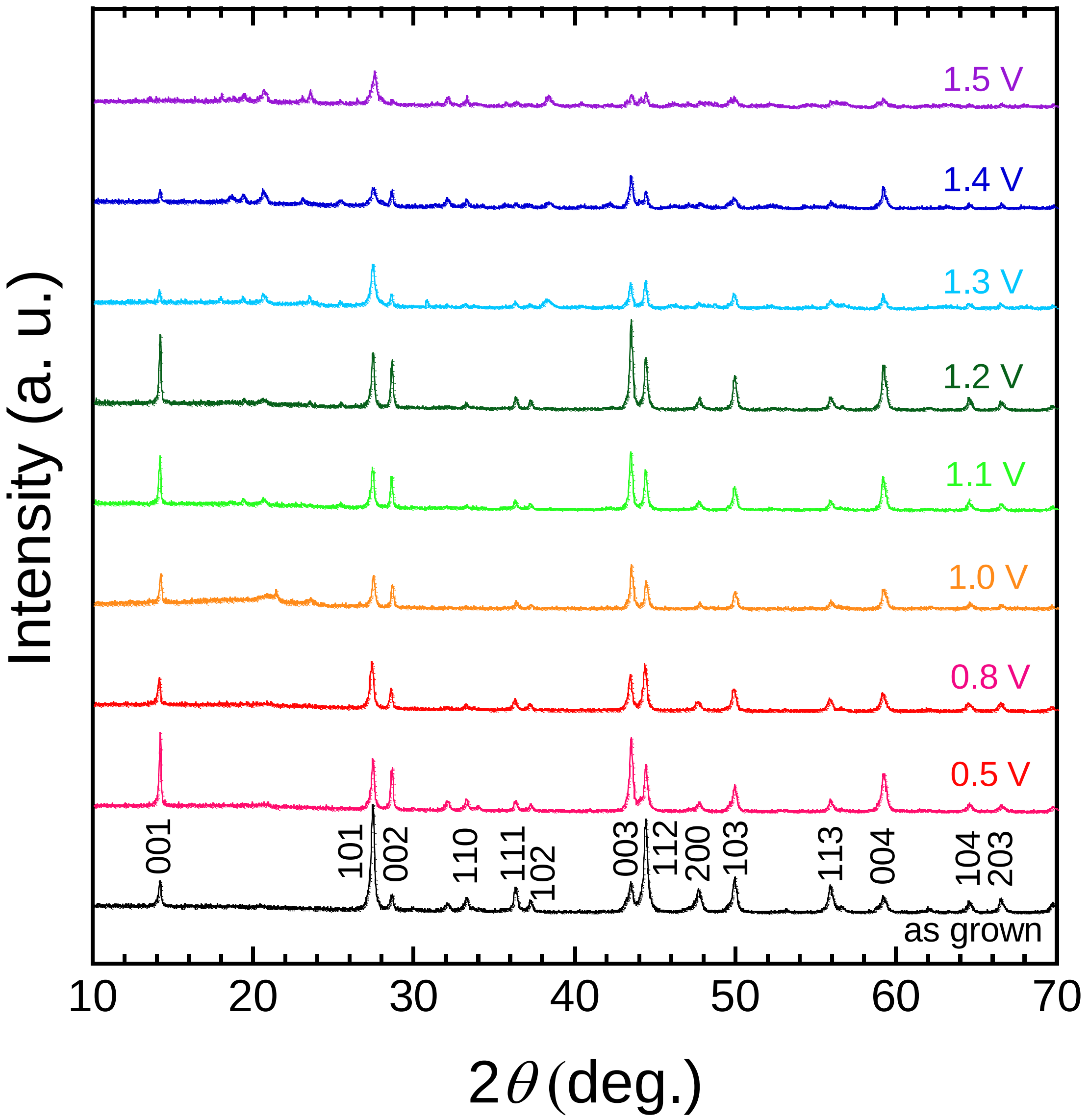

**Figure 1**

XRD patterns for the $FeTe_{0.8}S_{0.2}$ samples, as-grown and after the electrochemical reaction.

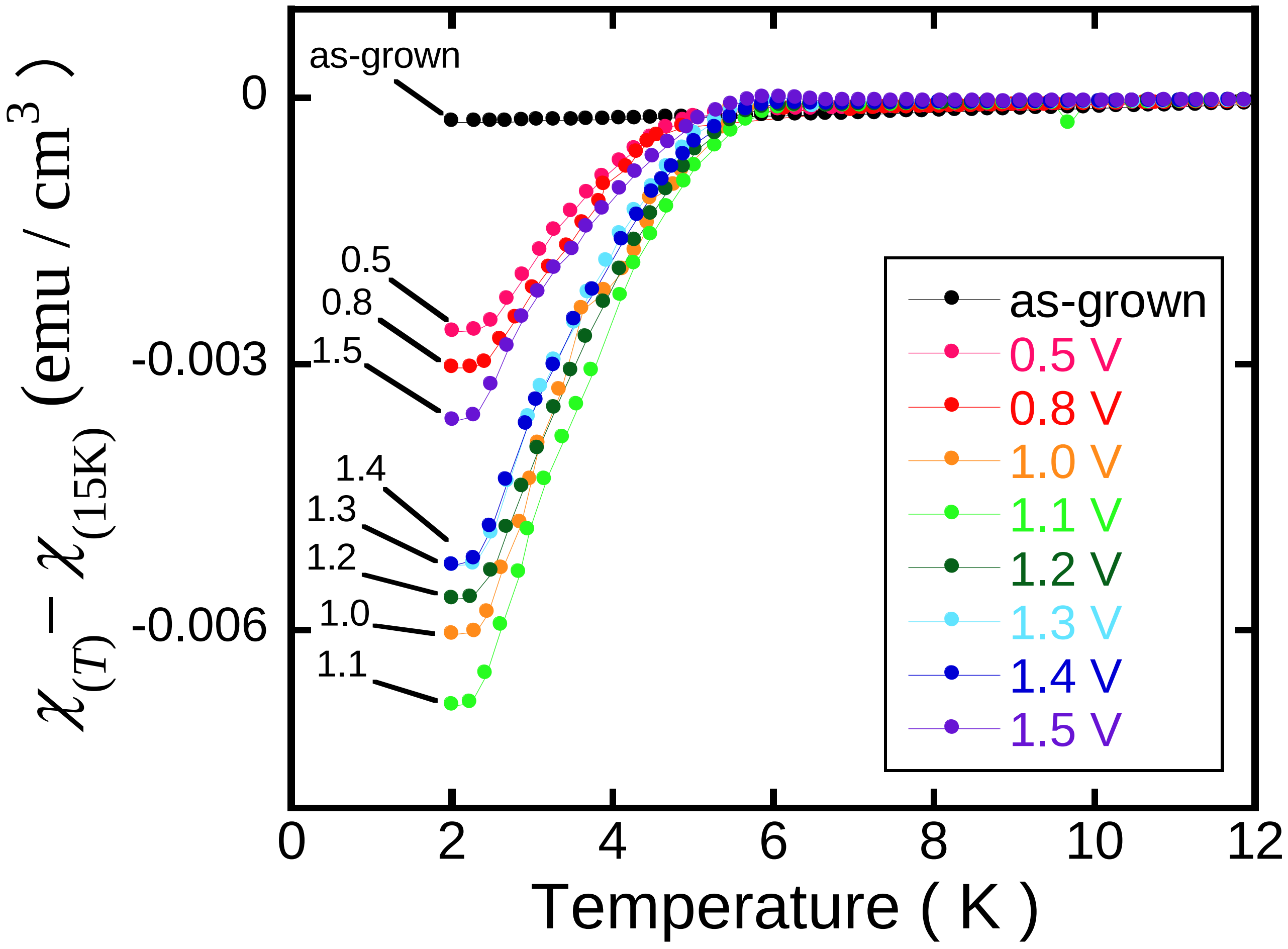

**Figure 2**

Temperature dependence of magnetic susceptibility for the samples obtained by the electrochemical reaction at various applied voltages. The susceptibility was normalized at a value of 15 K.

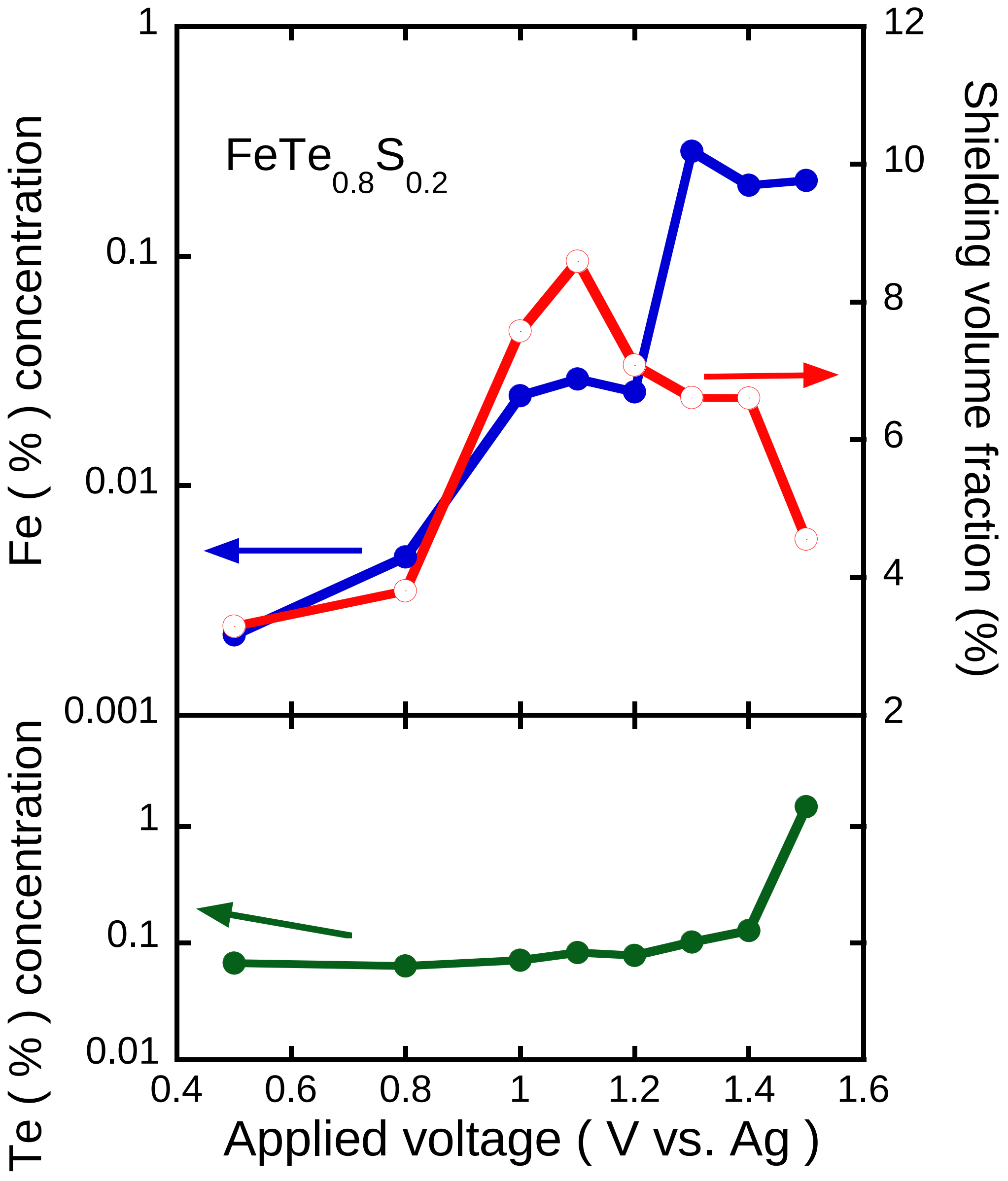

**Figure 3**

Applied voltage dependence of superconducting shielding volume fraction (right hand side of the figure) and Fe, Te concentration in the solution after the electrochemical reaction (left hand side of the figure) using a logarithmic scale.

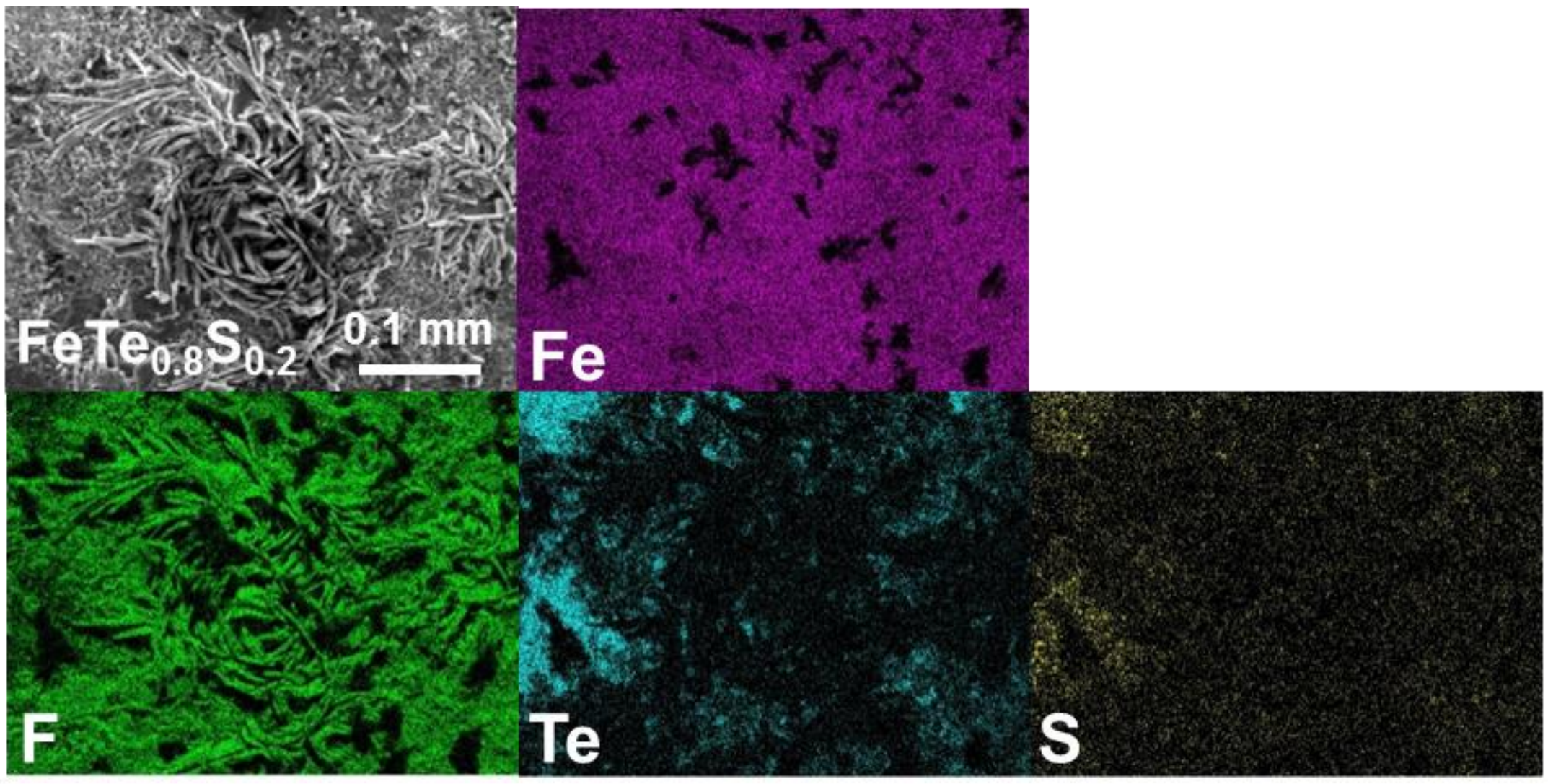

**Figure 4**

SEM images and EDX elemental mapping of $FeTe_{0.8}S_{0.2}$ sample after electrochemical reaction at 1.5 V.